# Implementation of ASK, FSK and PSK with BER vs. SNR comparison over AWGN channel

[1]Subrato Bharati, [2]Mohammad Atikur Rahman, [3]Prajoy Podder

[1,2,3]Department of EEE, Ranada Prasad Shaha University, Narayanganj, Bangladesh.

[3]Department of ECE, Khulna University of Engineering & Technology, Bangladesh.

[1]subratobharati1@gmail.com, [2]sajibextreme@gmail.com, [3]prajoypodder@gmail.com

*Abstract*— This paper mainly discusses about three basic digital modulation process ASK, FSK, PSK. These modulation schemes can be characterized by their transmitted symbols which consist of a discrete set of values occurring at gradually spaced intervals. The selection of a digital modulation technique for a specific application depend not only the bandwidth efficiency and implementation complexity but also error rate occurred in a bit (BER) and signal to noise ratio. Binary modulation methods use two level symbols and are facile to implement, provide good error substantiation. BER is a key parameter that used for assessing systems that transmit signal data from one location to another. SNR is well known measure of how the signal and noise power compare against each other. It directly affects the probability of error performance of a system. In this paper, we have implemented ASK, FSK and PSK using MATLAB. Cosine signal has been used as a carrier wave. In this paper, a comparative exploration of BER performance of ASK, FSK and PSK for channel utilization is proposed and the investigation are carried out with SNR over AWGN channel as the reference factor.

## I. INTRODUCTION

At present age is the day of communication. The biggest part of communication is growing by wireless technologies. The performance of the good transmitting and receiving systems is very important. For good performance, attenuation, distortion, noise must be avoided from transmitting signal as long as possible. For that measurement of transmitting signal and receiving signal should be accurate. With some digital modulation techniques (ASK, FSK, PSK), parameters, coding and filtering can affect the transmission quality and accuracy of the received signal. Digital modulation is the modulation techniques that are used to discrete signals to modulate a carrier wave. In digital modulation, high carrier frequencies are used so that signals can transmit over long distances with the help of criterion long distance communication media such as radio channel [1]. Noise in the channel does not have the deleterious effect on the received of demodulated signal is the main advantages. However, small, if an analog signal has some noise, the demodulated signal is corrupted. For example, if the modulating signal were in the range of 0 to 1 V, the specific value 0.58 V were sent, a small amount of noise might the change the value of demodulated signal 0.60 V, the receiver would believe the correct value was 0.60 V. The main advantage of ASK modulation for generation of ASK is that it's relatively easy to implement. It offers high bandwidth efficiency. Both ASK modulation and demodulation processes are relatively cheap. The ASK modulation technique is commonly used in transmitting digital data. In FSK, the implementation is easier than AM technique. PSK modulation has good noise rejection capability and the system generates a smaller noise bandwidth. Coding and modulation are the process of optimizing the performance of digital communication systems. In communication, parameters are used for the purpose of controlling error of communication system. BER and SNR parameters are used here for digital modulation technique.

Block Diagram of a Digital Communication System is shown in Fig.1.

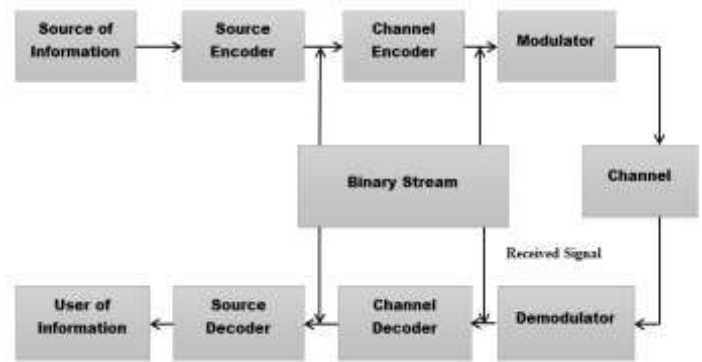

Fig.1.  Block Diagram of Digital Communication System

## II. THEORETICAL DESCRIPTION OF ASK, FSK AND PSK

### A. Amplitude Shift Keying (ASK)

Amplitude shift keying is a kind of Amplitude Modulation which represents the binary data (0 or 1) for variations in the amplitude of a carrier wave.

The amplitude shift keying is also called on-off keying (OOK). This is the simplest digital modulation technique. The amplitude of an analog carrier signal varies in accordance with the bit stream (modulating signal) where frequency and phase are keeping constant [2]. Any modulated signal has a high frequency carrier. The binary signal when ASK modulated, gives a zero value for low input and it gives the carrier output for high input. ASK can be expressed by,

$$S(t) = dA\cos\omega_c t \ldots\ldots\ldots(1)$$

We know that, $\omega_c = 2\pi f_c \ldots\ldots\ldots(2)$

or, $S(t) = dA\cos 2\pi f_c t \quad 0 \le t \le T \ldots(3)$

Here $f_c$ = frequency of the carrier wave and A= constant, d(t)= 1 or 0 , T = bit duration.

For d=0, S(t)= 0 and For d=1, S (t) = $A\cos 2\pi f_c t$……(4)

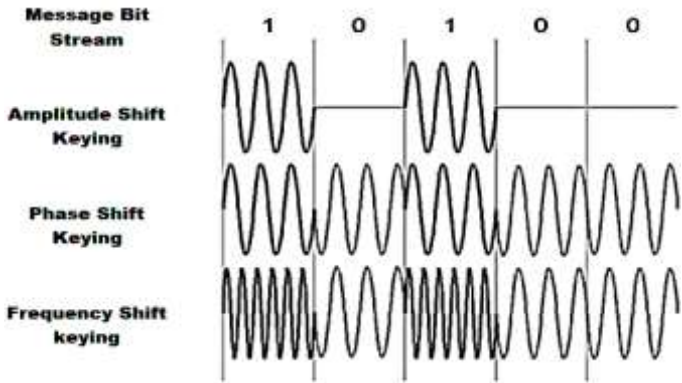

Fig.2. ASK, FSK and PSK Modulation

### B. Frequency Shift Keying(FSK)

Frequency Shift Keying is the method of transmitting digital signals in which the frequency of carrier signal varies according to the digital signal changes. FSK uses a pair of discrete frequencies to transmit binary (0s and 1s) information. With this scheme the one is called the mark frequency and the zero is called the space frequency. FSK is commonly used for caller ID and remote metering applications. It is also known as frequency modulation [3].

$$S_1(t) = A\cos(\omega_1 t + \theta_1) \quad ; \text{ for bit 1} \ldots\ldots(5)$$

$$S_2(t) = A\cos(\omega_2 t + \theta_2) \quad ; \text{ for bit 0} \ldots\ldots(6)$$

As, ω = 2πf So,

$$S_1(t) = A\cos(2\pi f_1 t + \theta_1) \quad ; \text{ bit 1} \ldots\ldots(7)$$

$$S_1(t) = A\cos(2\pi f_2 t + \theta_2) \quad ; \text{ bit 0} \ldots\ldots(8)$$

Here, A = Constant and f = frequency of the carrier wave

### C. Phase Shift Keying(PSK)

Phase shift keying is a digital modulation process which carries data by changing the phase of the carrier wave. There are several methods that can be used accomplish PSK. The binary phase shift keying technic is simpler then quadrature phase shift keying. In binary phase shift keying to opposite signal phases are used because there are two possible wave phase. The digital signal is separated time wise into individual bits. For the different bits phase will be changed for the two same values phase will be unchanged. Binary phase shift keying is sometimes called be phase modulation [8,9].

For the input binary sequence, binary input from 1 to 0 output modulated wave will change its phase at $180^0$. Binary input from 0 to 1 output modulated wave will change its phase at $180^0$. At the time of same binary input output modulated wave remain unchanged (from 0 to 0 or 1 to 1) or changed by 00. It is widely used for wireless LANs, RFID and Bluetooth communication. The problem with phase shift keying is that the receiver cannot know the exact phase of the transmitted signal to determine whether it is in a mark or space condition. This would not be possible even if the transmitter and receiver clocks were accurately linked because the path length would determine the exact phase of the received signal.

The binary phase shift keying signal can be defined by

$$S(t) = Am(t)\cos\omega t \quad ; 0 \le t \le T \ldots\ldots(9)$$

$$S(t) = Am(t)\cos 2\pi f_c t \quad ; 0 \le t \le T \ldots\ldots(10)$$

Here, m(t) = +1 or -1

A = Constant, $f_c$ = frequency of the carrier wave, T= bit duration

### D. Bit Error Rate (BER)

Bit error rate refers to the number of bit errors in per unit time. It is the ratio of total number of error bit to the total number of transmitted bit. It is very important way to determine the quality of transmission. It is often expressed as a percentage [4].

$$BER = \frac{Total\ number\ of\ error\ bits}{Total\ number\ of\ transmitted\ bits} \ldots\ldots(11)$$

As an example, let the transmitted bit sequence:
0 1 0 1 0 0 0 1 1 0
And the following received bit sequence:
0 1 <u>1</u> 1 0 <u>1</u> 0 <u>0</u> 1 <u>1</u>
In this case the number of bit errors is 4 (Underlined symbol).
$BER = \frac{4}{10} \times 100\%$
   $= 40\%\ error$

### E. Signal to Noise Ratio (SNR)

Signal to noise ratio means the ratio between the power of carrier signals to the power of noise signal in a wave. It is a measure used to compares the level of a desired signal to the level of background noise. It is expressed in logarithmic scale (dB). SNR also expressed by $\mu_0/2\mu$.

$$SNR = \frac{Signal\ Power}{Noise\ Power} \ldots\ldots\ldots\ldots (12)$$

### III. SIMULATION RESULTS

Fig.3 represents the simulation result of ASK modulation using MATLAB. Input bit Stream [1 0 1 1 0 1 0] and input signal frequency is 0.5 Hz. Cosine signal has been used as a carrier wave. Demodulation of ASK has also been performed to ensure the confirmation of proper modulation and demodulation.

Fig.4 shows the simulation result of FSK modulation where cos (10πft) and cos (2πft) are high frequency and low frequency carrier signal respectively. F and t are the frequency and time period respectively.

Fig.5 shows the simulation result of Phase shift keying modulation. It can be easily observed that Phase transition occurs in PSK modulated wave.

***BER vs. SNR:***   Error probabilities by communication theory can be calculated in terms of signal to noise ratio (SNR) [7]. There is inverse relation between signal to noise ratio (SNR) and bit error rate (BER). When BER increases the SNR decreases and When SNR increases the BER decreases. Generally, SNR Shows the quality of signal and BER show the error rate in signal. 10 $\log_{10}$ ($\mu_0/2\mu$) i.e. the mean photon number divided by $\mu=nsp$ and expressed in dB. It is indeed the signal to noise ratio SNR= $\mu_0/2\mu$ of the ASK receiver which must be considered [10].

In BER vs. SNR figure, we have found that both BER and SNR treated inversely as described above. For low SNR case, block performs better. Also for low BER case block performs better. Similarly, for high SNR case, block performs weaker. Also for high BER case block performs weaker. The BER and SNR of ASK and FSK treated almost same.

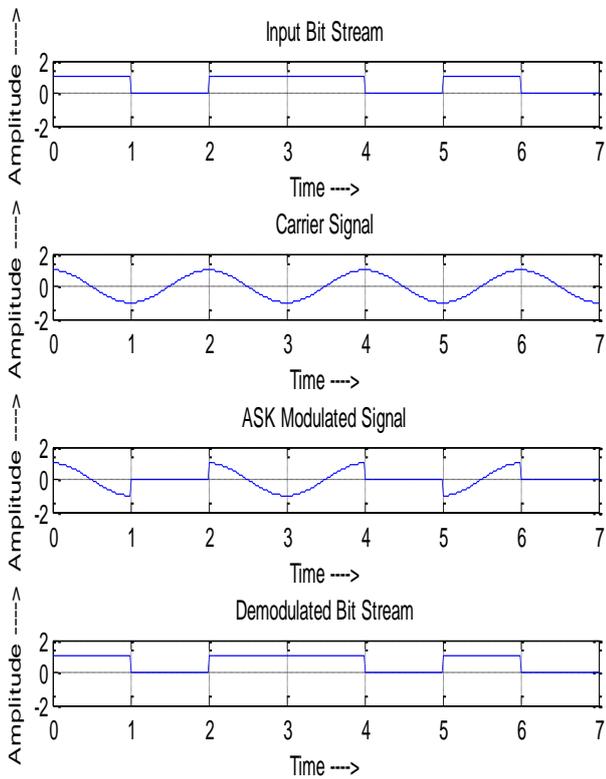

Fig.3.  Amplitude shift keying modulation and demodulation

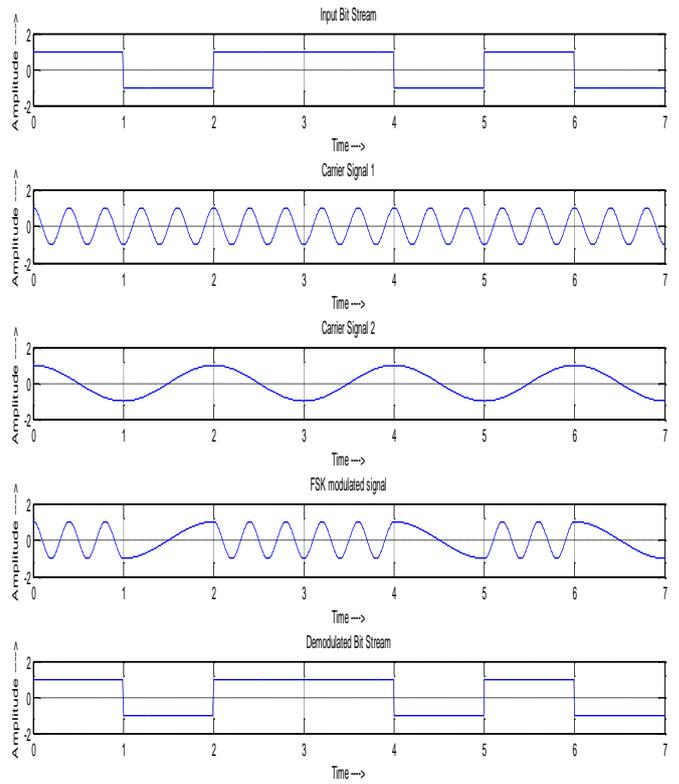

Fig.4.  Frequency shift keying modulation and demodulation

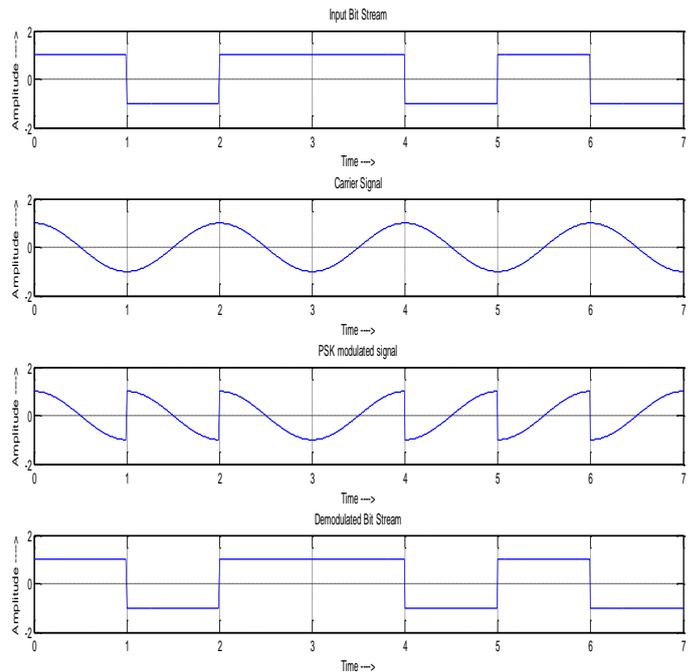

Fig.5.  Phase shift keying modulation and demodulation

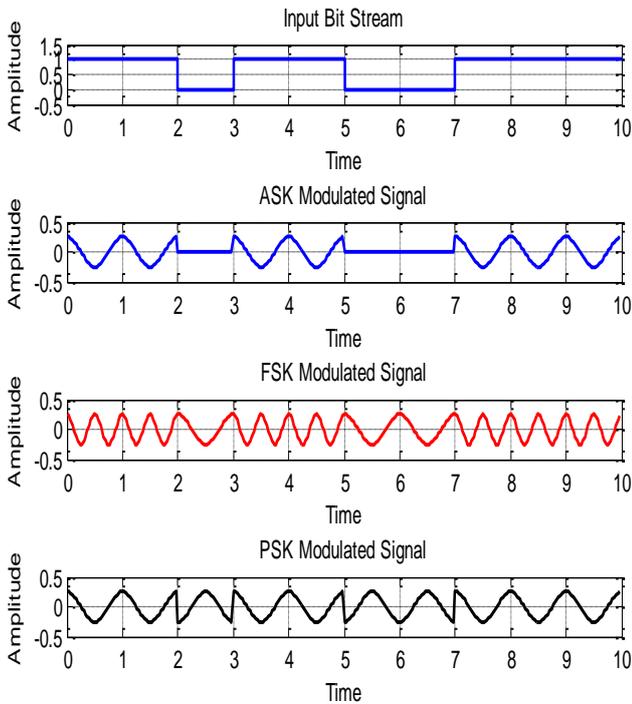

Fig.6. Input and modulated bit pattern when error rate is zero

Fig. 6 shows the ASK, FSK and PSK modulated signal with respect to input message signal and Fig. 7 indicates the BER vs. SNR curve. AWGN channel is considered for evaluating the BER performance. But original input Bit stream changes (Fig.8 ) compared to the input bit stream of fig. 6 when error rate is 0.2 and BER vs SNR characteristics (refer to fig.9) also changes for ASK, FSK and PSK..

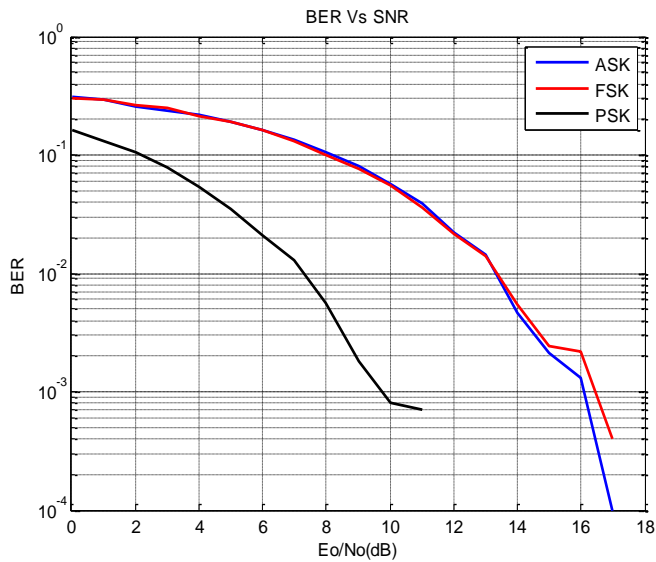

Fig.7. BER vs. SNR curve in the case of zero error rate

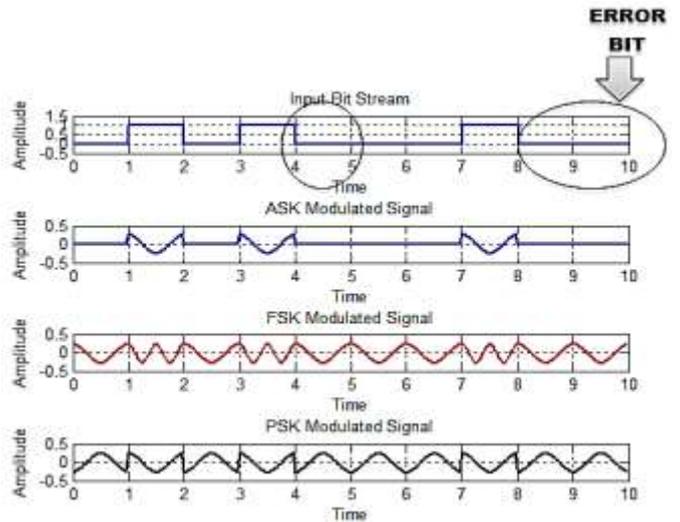

Fig.8. Input and modulated bit pattern when error rate is 0.2

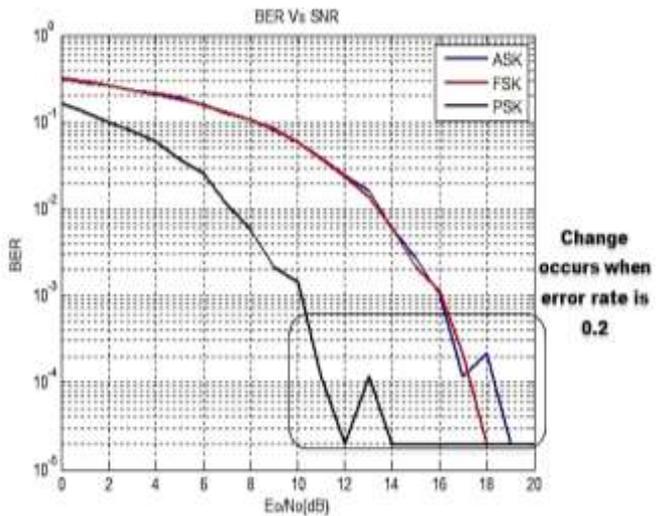

Fig.9. BER vs. SNR curve when error rate is 0.2

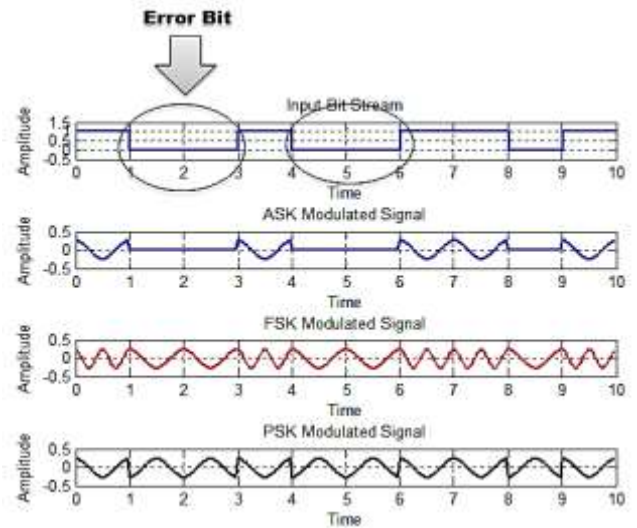

Fig.10. Input and modulated bit pattern when error rate is 0.4

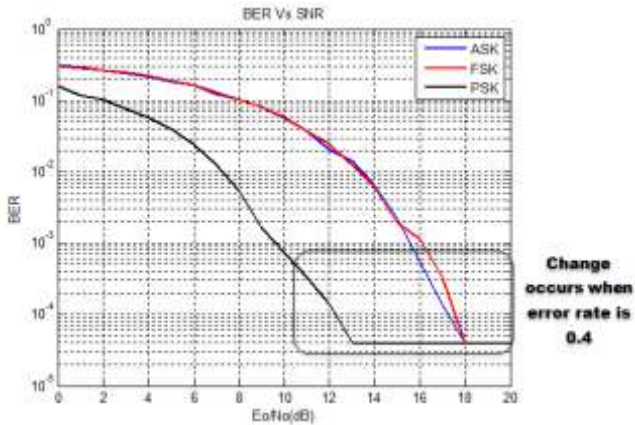

Fig.11. BER vs. SNR curve when error rate is 0.4

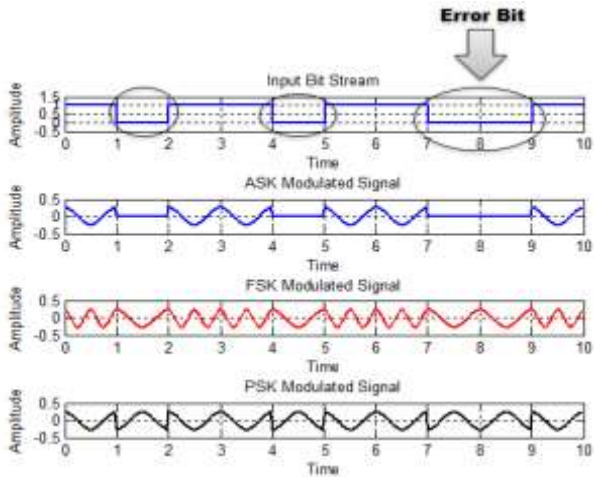

Fig.12. Input and modulated bit pattern when error rate is 0.6.

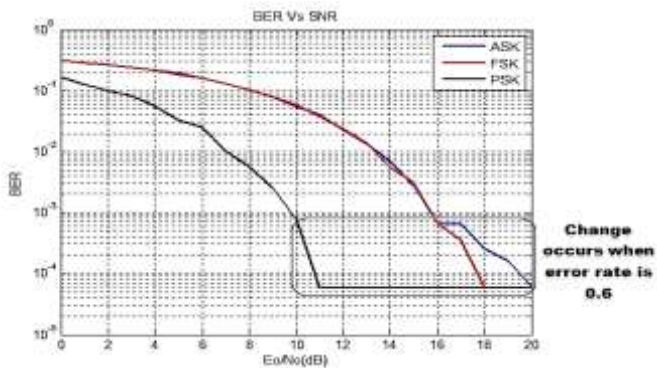

Fig.13. BER vs. SNR curve when error rate is 0.6

Fig. 10 and 12 describes how the original bit stream changes and error bit increase when error rate is 0.4 and 0.6 respectively. Fig. 11 and 13 explicates the BER vs. SNR comparison curve for error rate 0.4 and 0.6 respectively.

## IV. CONCLUSION

In this paper, three digital modulation schemes have been discussed theoretically and then implemented them in MATLAB simulation. We have also demodulate the modulated ASK, FSK and PSK and recovered the original bit stream or message signal. Then the performance (BER vs. Eo/No) of ASK, FSK and PSK over Additive white Gaussian noise (AWGN) channel has been observed when bit error rate is zero and when the bit error rate has a high value. The position of error bit in input bit stream has also been localized and mentioned in the simulation result in the case of error rate.

REFERENCES

[1] William L. Schweber, "Data Communications," McGraw-Hill Book Company, 1988.

[2] J.S.Chitode, "Digital Communications," Technical Publications, 2009.

[3] R. P. Singh, S. D. Sapre, "Communication Systems," 2nd Edition, Tata McGraw-Hill Education, 2008.

[4] M. Muktadir Rahman, Prajoy Podder, Tanvir Zaman Khan, Mamdudul Haque Khan, "BER Performance Analysis Of OFDM-BPSK, QPSK, QAM Over Rayleigh Fading Channel & AWGN Channel," International Journal of Industrial Electronics and Electrical Engineering(IJIEEE), Vol.2, Issue. 7, pp. 1-5, July 2014.

[5] Roger L. Freeman, "Telecommunication System Engineering," John Wiley & Sons, 2004.

[6] Lathi, B.P., "Modern Digital and Analog Communication," Oxford University Press, New York, 1998.

[7] D. Dutt Bohra and A. Bora, "Bit Error Rate Analysis in Simulation of Digital communication Systems with different Modulation Schemes", International Journal of Innovative Science, vol. 1, no. 3, 2014.

[8] Simon S. Haykin, Michael Moher, "Introduction to analog and digital communications," Wiley Publisher, 2007.

[9] David R. Smith, "Digital Transmission Systems," Springer Science & Business Media, 2003.

[10] Md. Mehedi Hasan, Jag Mohan Thakur, Prajoy Podder, "Design & Implementation of FHSS and DSSS for Secure Data Transmission,"International Journal of Signal Processing systems, Vol.4, No.2, pp.144-149, April 2016.